\documentclass[11pt,aaspp4]{aastex}

\newcommand\CQ{core quad}
\newcommand\LQ{long-axis quad}
\newcommand\SQ{short-axis quad}
\newcommand\IQ{inclined quad}
\newcommand\LD{long-axis double}
\newcommand\SD{short-axis double}
\newcommand\AD{axial double}
\newcommand\ID{inclined double}

\newcommand\1{{\tt 1}}
\newcommand\2{{\tt 2}}
\newcommand\3{{\tt 3}}
\newcommand\4{{\tt 4}}

\def\half{\hbox{$1\over2$}}
\def\hoverarrow#1{\setbox0\hbox to 0pt{\hss$\scriptstyle\rightarrow$}%
                  #1\kern.6ex\raise 1.6ex\box0\kern-.2ex}
\def\(#1){\raise.3ex\hbox{(}#1\mskip.5\thinmuskip\raise.3ex\hbox{)}}
\def\btheta{{\hoverarrow\theta}}
\def\bthim{\btheta_{\rm im}}
\def\bbeta{{\hoverarrow\beta}}
\def\zl{z_{\rm L}}

\def\witchbox#1#2#3{\hbox{$\mathchar"#1#2#3$}}
\def\leqsim{\mathrel{\rlap{\lower3pt\witchbox218}\raise2pt\witchbox13C}}
\def\geqsim{\mathrel{\rlap{\lower3pt\witchbox218}\raise2pt\witchbox13E}}
\def\bquote{\smallskip\bgroup\leftskip=35pt\rightskip=34pt
            \frenchspacing\em}
\def\equote{\smallskip\egroup}

\begin{document}


\shortauthors{P. Saha \& L.L.R. Williams}

\shorttitle{Lensed QSOs}

\title{Qualitative Theory for Lensed QSOs}

\author{Prasenjit Saha}
\affil{Astronomy Unit, School of Mathematical Sciences\\
       Queen Mary and Westfield College\\
       University of London\\
       London E1~4NS, UK}
\and
\author{Liliya L.R. Williams}
\affil{Department of Physics and Astronomy\\
       University of Minnesota\\
       116 Church Street SE\\
       Minneapolis, MN 55455}

\begin{abstract} 
We show that some characteristics of multiply-imaged QSO systems are
very model-independent and can be deduced accurately by simply
scrutinizing the relative positions of images and galaxy-lens center.
These include the time-ordering of the images, the orientation of the
lens potential, and the rough morphology of any ring.  Other features
can differ considerably between specific models; $H_0$ is an example.
Surprisingly, properties inherited from a circularly symmetric lens
system are model-dependent, whereas features that arise from the
breaking of circular symmetry are model-independent.  We first develop
these results from some abstract geometrical ideas, then illustrate
them for some well-known systems (the quads Q2237+030, H1413+117,
HST14113+5211, PG1115+080, MG0414+0534, B1608+656, B1422+231, and
RXJ0911+0551, and the ten-image system B1933+507), and finally remark
on two systems (B1359+154 and PMN J0134-0931) where the lens
properties are more complex.  We also introduce a Java applet which
produces simple lens systems, and helps further illustrate the
concepts.
\end{abstract}


\section{Introduction}\label{intro}

Multiply-imaged QSOs are studied in pursuit of a variety of goals.
The positions and relative magnifications of the images reveal
something about the mass distribution of lensing galaxies, and if
microlensing is observable then they also probe the mass function on
stellar scales. If time delays are measurable they supply estimates of
the Hubble constant $H_0$, and image statistics over many different
lenses indicate something about the dimensionless global cosmological
parameters, $\Omega$ and $\Lambda$.  Microlensing also constrains the
intrinsic sizes of the sources, and possibly the bulk velocity or
velocity dispersion of the galaxy.  Finally, individual images in a
multiple-image system provide nearby but distinct lines of sight
through the intergalactic medium.  These are discussed at length in
recent reviews \citep{w98,cs02,courb03} and older but still relevant
works \citep{w90,sef92,bn92,n98}.

With so much interest, it is not surprising that many models of galaxy
lenses have been developed, employing several different methods.  
What is somewhat surprising, however, is that some features of 
individual lenses are robustly model-independent while others are very
model-dependent.  We defer detailed references until the section on
case studies, but for now we remark that different models of any given
system always seem to agree about four aspects:
\begin{enumerate}
\item the time-ordering of the images, i.e., the ordering of time
delays if the source is variable (but not the {\it values\/} of the
time delays);
\item the direction of the source's displacement, relative to the lens
center;
\item the orientation of the lens-potential's ellipticity, if
significant; such ellipticity often indicates external shear from
other galaxies;
\item the rough morphology of an Einstein ring, which can be predicted
from QSO images alone.
\end{enumerate}
At the same time, a much sought-after quantity:
\begin{enumerate}
\item[0.] the projected radial density profile, which decides the mass 
normalization and the conversion between time delays and $H_0$,
\end{enumerate}
is highly model-dependent.

Since the points 1--4 are model-independent, one may wonder if they 
can be understood in a model-independent way. In this paper we show that
indeed they can, using some abstract but basically geometrical ideas.
The key concepts were introduced by \citet{bn86}, who really initiated
the qualitative theory of lensing. Other aspects of this paper were
anticipated by \citet{bk88}, and \citet{wucknitz02}.

\section{Some invisible curves}

Like many areas of physics, gravitational lensing can be elegantly
approached through a variational principle, Fermat's principle,
which has been formulated for gravitational lensing through
several routes \citep{schneider85,bn86,kov90,ns92}.  Of particular
interest is the arrival-time surface $\tau(\btheta)$, which is the
light-travel time (in suitable units) for a virtual photon arriving
from sky-position $\btheta$, visualized as a surface.  According to
Fermat's principle, real photons come from $\btheta$ for which
$\tau(\btheta)$ is stationary; that is, images form where the
arrival-time surface has a maximum, minimum, or saddle point.

The arrival-time surface is a function of the unlensed sky-position
$\bbeta$ and the sky-projected density (again, in suitable units)
$\kappa(\btheta)$. A concise expression for the arrival time is
\begin{equation}
\tau\(\btheta) = \half\(\btheta-\bbeta)^2 - 2\nabla^{-2}\kappa\(\btheta) .
\label{arriv-eq}
\end{equation}
The first term in Equation (\ref{arriv-eq}) is geometrical, while the
second term is the gravitational time delay due to the lens, also called
the lens potential and written as $\psi(\btheta)$. The unusual symbol
$\nabla^{-2}$ is just an inverse Laplacian operator; it denotes the
operation of solving Poisson's equation (in two dimensions).

For calculations it is conventional to use units such that $\kappa$ is
dimensionless and of order unity while $\tau$ is also dimensionless
and measured in arcsec$^2$.  But the arrival-time $t(\btheta)$ and
surface $\Sigma(\btheta)$ in physical units are easily recovered
through the relations
\begin{eqnarray}
  & & t(\btheta) = T_0 \times \tau(\btheta), \qquad
    \Sigma(\btheta) = \Sigma_0 \times \kappa(\btheta) , \\
  \noalign{\medskip}
  & & T_0 = \displaystyle{ h^{-1} {(1+\zl)\over c}
                           {D_{\rm L}D_{\rm S}\over D_{\rm LS}} }
    \simeq \zl \times {100 \, h^{-1} \rm \,days\,arcsec^{-2}} ,
    \label{eqT0} \\
  \noalign{\medskip}
  & & \Sigma_0 = \displaystyle{ {c^3\over 4\pi G}
                                {T_0\over(1+\zl)} }
    \simeq \zl \times {10^{11}\,h^{-1}\,M_\odot\rm\,arcsec^{-2}} .
\end{eqnarray}
There is really only one scale in the problem: $T_0$, which is of
order the Hubble time; the density scale $\Sigma_0$ (known as the
critical density) is essentially equivalent.

The derivatives of the arrival time are very important.  The condition
for an image at $\btheta=\bthim$, being that the arrival time has a
maximum, minimum, or saddle point at $\bthim$, is
\begin{equation}
\nabla\tau(\bthim) = 0.
\label{lens-eq}
\end{equation}
Equation (\ref{lens-eq}) defines a mapping from $\bthim$ to $\bbeta$,
which is the lens equation.  Further, $\nabla\nabla\tau\(\bthim)$ is
the inverse (tensor) magnification of the image.

Elsewhere we have developed modeling techniques that concentrate on
reconstructing the arrival-time surface.  At first we thought of
discretizing $\tau$ \citep{ws95} but later concluded \citep{sw97} that
discretizing $\kappa$ is a better idea.  In the present paper we are
interested not in model-making as such, but in gaining intuition for
the shape of the arrival-time surface, the locations of images, and 
how the two relate to each other. For this purpose, it is useful to 
introduce some abstract (and hence invisible) curves.

\subsection{Saddle point contours}

Saddle-point contours are special contours in the arrival-time
surface: contours that pass through saddle points, or equivalently,
self-crossing contours, as first discussed in connection with
gravitational lensing by \citet{bn86}.  
Their usefulness is that they form a sort of skeleton for a 
multiple-image system, as we now explain.

\begin{figure}
\epsscale{.7}
\plotone{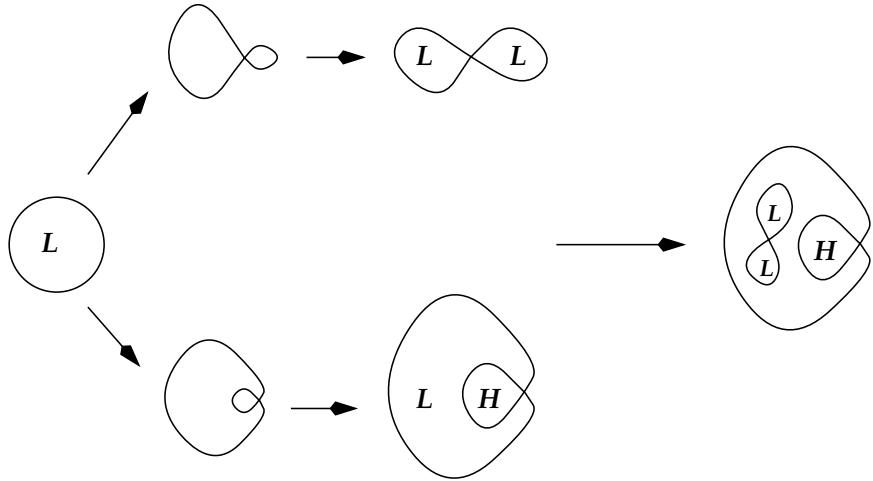}
\caption{Multiple images via saddle-point contours in the arrival-time
surface.  Here $L$ marks minima and $H$ marks maxima.  (Figure by
H.M. AbdelSalam.)}
\label{fig-lemlim}
\end{figure}

In the absence of lensing $\tau\(\btheta)$ is a parabola, 
all the contours are circular, and the sole
image is at the minimum $\btheta=\bbeta$.
For a small lensing mass,
the shape changes slightly from being a parabola and the minimum moves
a little.  For large enough mass, a qualitative change occurs, in that
a contour becomes self-crossing.  There are two ways in which a
self-crossing can develop: as a kink on the outside of a contour line
(a figure `8', called a lemniscate), or a kink on the inside (called a
lima\c con).  These are illustrated in Figure~\ref{fig-lemlim}. With
the original contour having enclosed a minimum, a lemniscate produces
another minimum, plus a saddle-point at the self-crossing, while a
lima\c con produces a new maximum plus a saddle point.  (The previous
sentence remains valid if we interchange the words `maximum' and
`minimum'.)  The process of contour self-crossing can then repeat
around any of the new maxima and minima, producing more and more new
images, but always satisfying \begin{equation} \hbox{maxima} +
\hbox{minima} = \hbox{saddle points} + 1.  \end{equation} The
self-crossing or saddle-point contours form a sort of skeleton for the
multiple-image system.  Many different configurations are possible in
principle and are tabulated in \citet{bn86}, but in practice lensed
quasars characteristically have one of two configurations: double
quasars have a lima\c con, while quads have a lemniscate inside a
lima\c con, as in the rightmost part of Figure \ref{fig-lemlim}.  A
lima\c con is needed to produce a maximum (marked `{\it H\/}' in the
Figure \ref{fig-lemlim}) which will be located at the center of the
lensing galaxy. Since galaxies tend to have sharply-peaked central
densities, the corresponding image is highly demagnified and (so far)
always unobservable.\footnote{However, a central image is seen in some
clusters, notably in Cl0024+1654 (see, for example, Astronomy Picture
of the Day, {\tt http://antwrp.gsfc.nasa.gov/apod/ap980614.html}).}
Thus lensed quasars are doubles or quads: an incipient third or fifth
image hides at the center of the lensing galaxy.\footnote{In cluster
lenses, however, a single density peak does not always dominate, and
other image configurations are possible.  Three-image configurations
involving an isolated lemniscate are common, and a lima\c con with the
maximum observable is also possible.  See \cite{asw98} for examples in
Abell~370.}

\subsection{Critical curves and caustics}

Critical curves and caustics were introduced into gravitational
lensing by \cite{bk75}, and later explored more fully by \cite{bn86}.  

They involve derivatives of the arrival time,
and thus are not as intuitive as saddle-point contours. But they are
perhaps better known.

Critical curves are curves on the image plane where the magnification
is infinite.  More formally, they are curves where
$\nabla\nabla\tau\(\btheta)$ has a zero eigenvalue.  Clearly, at
minima of $\tau$ both eigenvalues of $\nabla\nabla\tau$ will be
positive, at maxima both eigenvalues will be negative, and at saddle
points one eigenvalue will be positive and one negative.  Thus
critical curves separate regions of the image plane that allow minima,
saddle points, and maxima.

Mapping critical curves from the $\btheta$ to $\bbeta$ via the lens
equation (\ref{lens-eq}) gives caustic curves.  Caustics separate
regions on the source plane that give rise to different numbers of
images.

In non-astrophysical optics, caustics are curves not on the source
plane but on the `observer plane'.  For example, the patterns at the
bottom of a swimming pool are caustics.  In fact, caustics can be
defined both on the source plane and the observer plane---after all,
light rays can run forwards or backwards.  In gravitational lensing we
are more interested in caustics on the source plane because we are
(usually) concerned with one observer on the ground and many sources
on the sky, whereas with a swimming pool the reverse applies.

\subsection{Apparent-shear ellipse}\label{appshear}

The apparent-shear ellipse is the minimum-area ellipse passing through
the four images of a quad.  Being specific to quads, the apparent-shear
ellipse does not have the same deep significance as the other
invisible curves, but is still very interesting. It was introduced by
\cite{wucknitz02} on the basis of models.  Here we will motivate it
in a model-independent way.

The geometrical part of the arrival time in equation~(\ref{arriv-eq})
is constant on circles.  The mass distributions $\kappa\(\btheta)$ of
galaxies do have significant ellipticity, but $\nabla^{-2}$ tends to
suppress the ellipticity in the potential. As a result, the images in
a quad produced by a galaxy potential will be roughly in circles
centered at $\bbeta$.  The situation changes if other group galaxies
or a cluster make a large contribution to the potential, because then
the potential can become more elliptical. In particular, consider a
constant external shear, given by a potential
$\half\gamma_1(\theta_x^2-\theta_y^2)+\gamma_2\theta_x\theta_y$.
The arrival-time will be
\begin{equation}
\tau\(\btheta) = \half\(\btheta-\bbeta)^2
+\half\gamma_1(\theta_x^2-\theta_y^2)+\gamma_2\theta_x\theta_y
- 2\nabla^{-2}\kappa_{\rm galaxy}\(\btheta) .
\label{ecc-ellip}
\end{equation}
Here the geometrical plus shear terms taken together behave like a
single geometrical term that is constant on ellipses of eccentricity
$\gamma=\sqrt{\gamma_1^2+\gamma_2^2}$.  If the galaxy potential is
fairly circular the images of a quad will still lie on an eccentric
ellipse.

Motivated by the above arguments, we consider ellipses
\begin{equation}
\half\(\btheta-\bbeta')^2
+\half\gamma_1'(\theta_x^2-\theta_y^2)+\gamma_2'\theta_x\theta_y
= c
\end{equation}
and solve for $\beta_x',\beta_y',\gamma_1',\gamma_2',c$ such that
(i)~all four images of a given quad lie on the ellipse, and
(ii)~$\gamma_1'^2+\gamma_2'^2$ is minimized.  (This is just a matter
of numerically solving 5 simultaneous linear equations.)  We call the
fitted value of $\sqrt{\gamma_1'^2+\gamma_2'^2}$ the `apparent shear'
or $\gamma_{\rm app}$. (Wucknitz calls it the `critical shear' because
in his models it has an interesting relation to the critical density.)
The apparent shear is not the actual shear
$\gamma=\sqrt{\gamma_1^2+\gamma_2^2}$, because it ignores dependence
of image positions on $\kappa_{\rm galaxy}\(\btheta)$.  But still, we
expect the apparent shear will give some indication of the direction
and scale of the true shear.

\section{Model-independent features}\label{topo}

\begin{figure}
\epsscale{.55}
\plotone{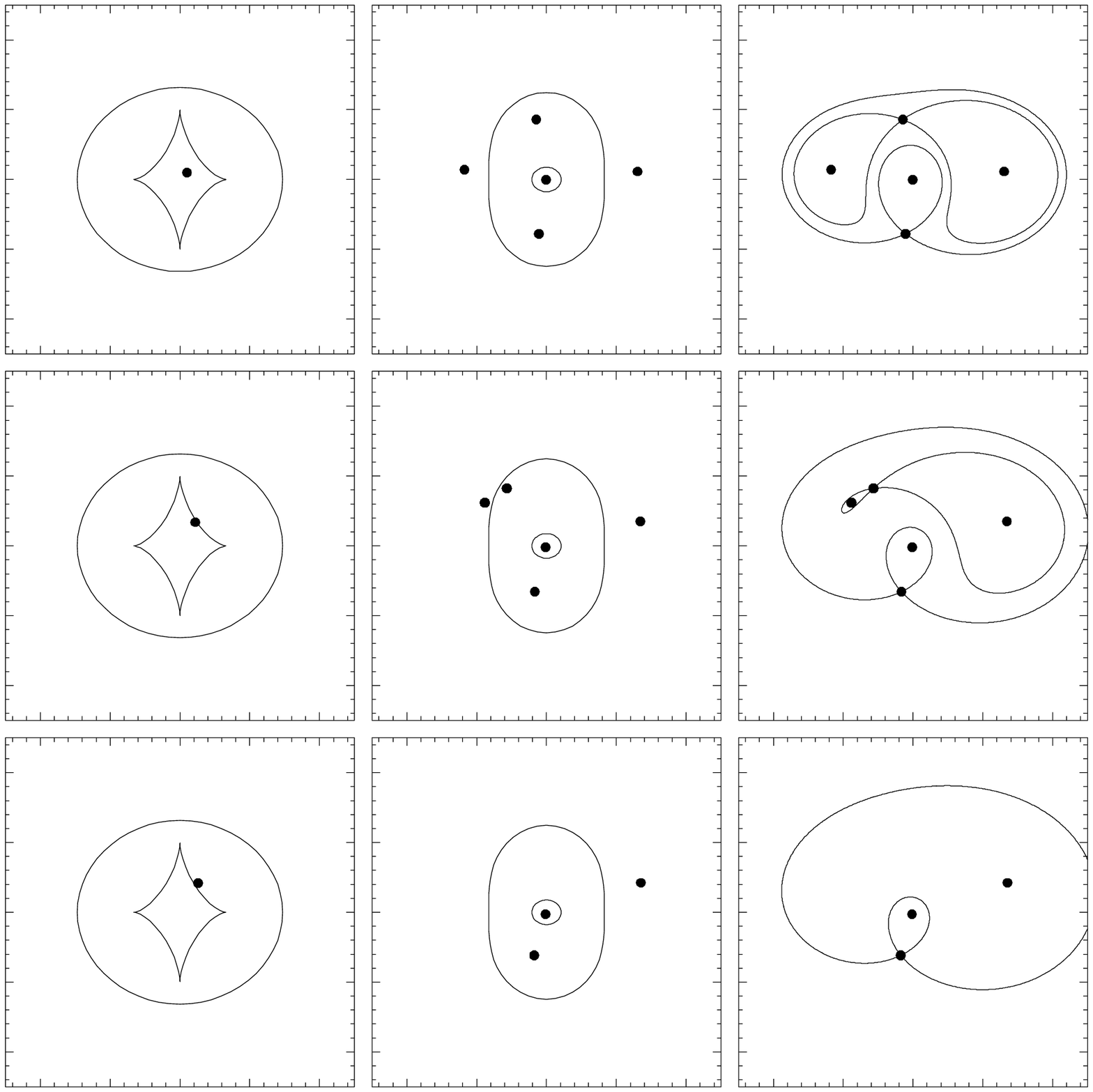}
\caption{Nonsingular isothermal galaxy-lens with external shear along the 
vertical axis.  Upper panels show a 
\CQ, middle panels show an \IQ, lower panels show an \ID. Left panels show
caustics in the source plane (expanded by a factor of 2) and the source, 
central panels show critical lines and images in the lens plane, 
and right panels show saddle point contours and images.}
\label{fig-inc}
\end{figure}

\begin{figure}
\epsscale{.55}
\plotone{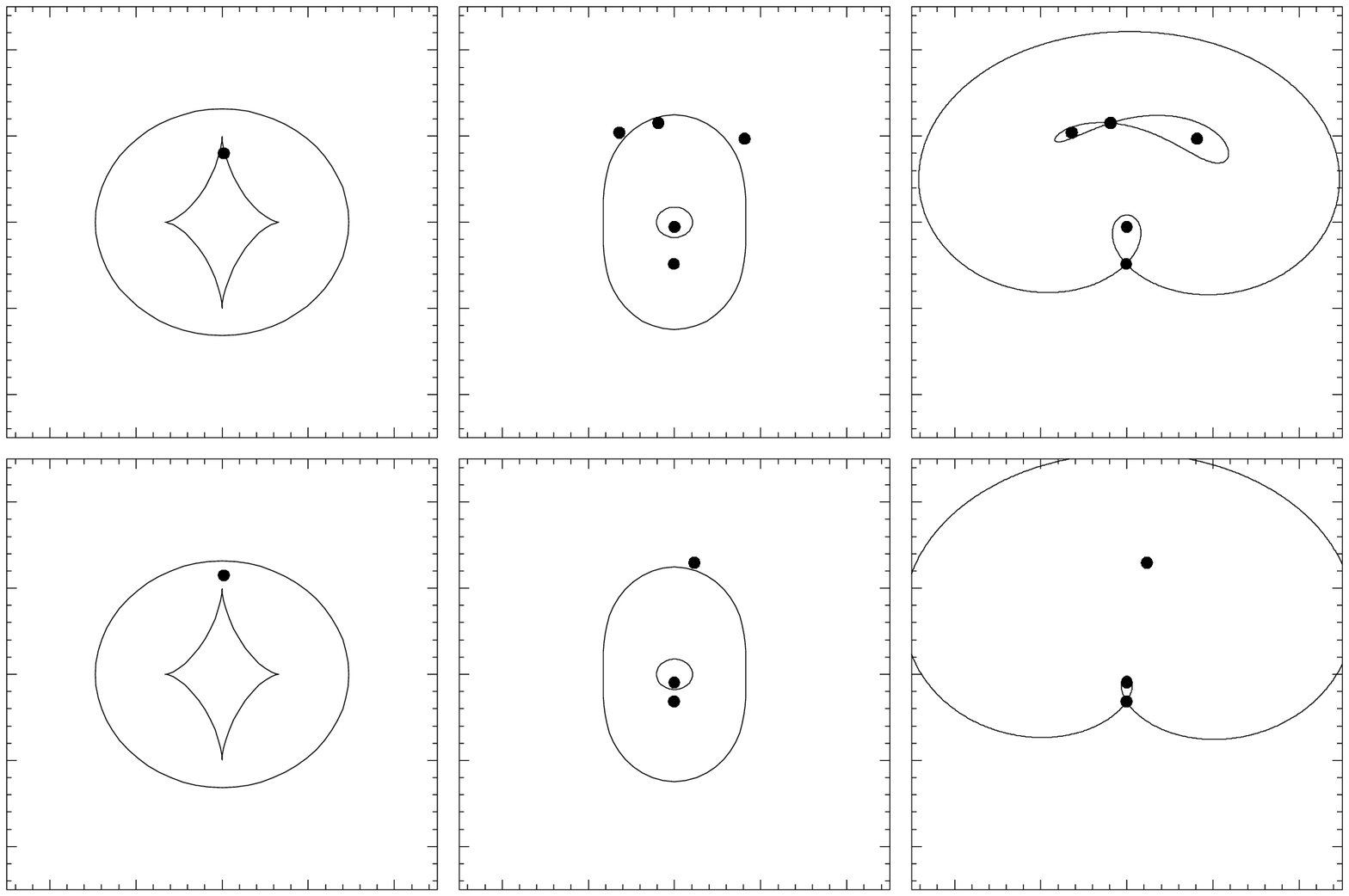}
\caption{Same lens as in Figure\ \ref{fig-inc}. Upper panels show a \LQ\
and lower panels show an \AD, in this case a \LD.}
\label{fig-long}
\end{figure}

\begin{figure}
\epsscale{.55}
\plotone{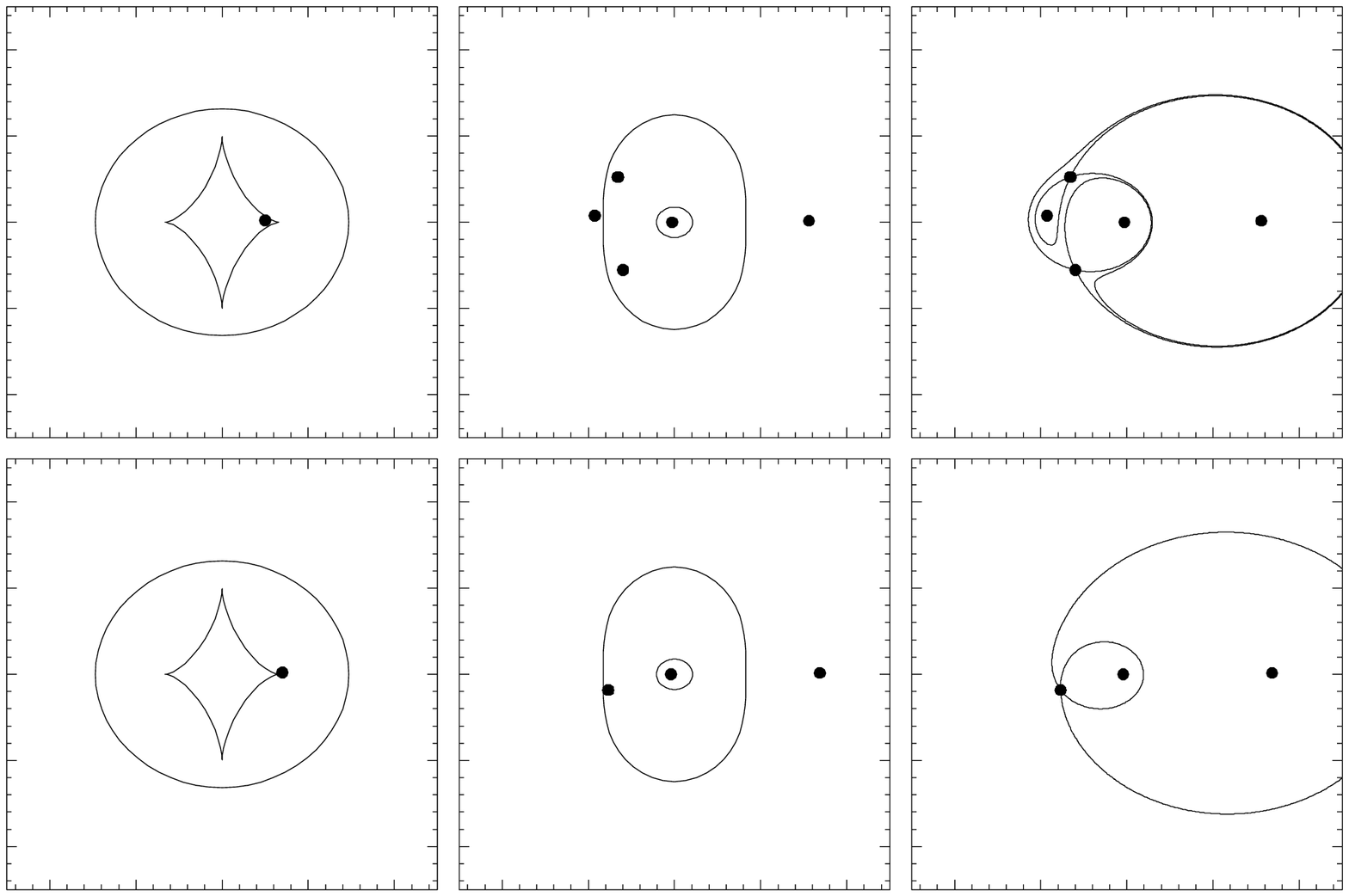}
\caption{Same lens as in Figures~\ref{fig-inc} and \ref{fig-long}.
Upper panels show a \SQ\ and lower panels show an \AD, in this case a \SD.}
\label{fig-short}
\end{figure}

We now study the qualitative features of lensing by typical galaxy
potentials: elliptical, but having a high central density.  A simple
example of such a potential is
\begin{equation}
\psi(\btheta) = (\theta_x^2+\theta_y^2+r_{\rm c}^2)^{\frac12} +
    \half\gamma(\theta_x^2-\theta_y^2)
\label{nsixs-eq}
\end{equation}
which is a non-singular isothermal with core radius $r_{\rm c}$ plus
external shear $\gamma$ corresponding to external mass to the N or S.
In Figures \ref{fig-inc} to \ref{fig-short}\footnote{Figures of this
type, i.e., showing source and image positions in source and lens
plane, with caustics and critical curves, respectively, were first
used by \citet{ng89} and \citet{bkkn89}.}  we take $r_{\rm c}=0.1$ and
$\gamma=0.2$ in the potential (\ref{nsixs-eq}) as an illustration, and
then examine what happens for different source positions, through
caustics, critical curves, and saddle-point contours.  Of course, the
features we will discuss do not depend on the details of the
potential.

The layout in Figures \ref{fig-inc} to \ref{fig-short} is as
follows.\footnote{An interactive version of Figures \ref{fig-inc} to
\ref{fig-short}, as a Java applet, is accessible via \hfil\break
{\tt http://ankh-morpork.maths.qmul.ac.uk/\char'176saha/lens/}}
\begin{itemize}
\item The right panels show saddle-point contours and images.  The
central image is always a maximum, the contour self-crossings are
saddle points, while the remaining images are minima of the arrival
time.
\item The middle panels show critical curves and images.  The outer
critical line is aligned with the long axis of the potential.  Recall
that critical lines are where an eigenvalue of the magnification
changes sign.  The consequence for this lens is that any image outside
both critical curves is a minimum, any image between the critical
curves is a saddle point, and any image inside both critical curves is
a maximum, all irrespective of the source position. Image parities can
be verified by comparing with the saddle-point contours.
\item The left panels show caustics and source positions.  (For better
visibility, the caustic panels have been expanded by a factor of two
relative to the other panels.) The two caustic curves demarcate
regions from where a source produces 1, 3, and 5 images.  The two
critical lines are maps of the caustics to the image plane, but the
{\it inner\/} (diamond) caustic maps to the {\it outer\/} critical
line.
\end{itemize}

\subsection{A classification scheme for doubles and quads}

Depending on the position of the source relative to the diamond
caustic of an elliptical lens we can identify four kinds of quads and
two kinds of doubles. We call these {\it\CQ s,} {\it\IQ s,} {\it\LQ s,} and
{\it\SQ s,} {\it\ID s,} and {\it\AD s}.

The top panels in Figure \ref{fig-inc} show a \CQ: the source is near
the center of the diamond caustic, and the images are distributed in a
cross.  (We disregard the unobservable central image.)  The middle
panels in the figure shows an \IQ: the source is just inside the
diamond caustic and away from the cusps; two of the images (a saddle
point and a minimum) are close together, with the remaining two
roughly 120$^\circ$ on either side.  The bottoms panels show an \ID:
the source has moved outside the diamond caustic, and two images have
merged and disappeared.

The upper panels in Figure \ref{fig-long} show a \LQ: the source is
just inside a cusp in the direction of the long axis of the lens
potential; one image (a saddle point) is close to the center while
the other three are on the opposite side and further out.  The lower
panels show an \AD: the source has moved outside the diamond caustic
and two images have merged and disappeared, leaving two images nearly
colinear with the center of the lens.

The upper panels in Figure \ref{fig-short} show a \SQ: here the source
is just inside a cusp in the direction of the short axis of the lens;
three images are close to the galaxy while one (a minimum) is on the
opposite side and further out. The lower panels show another \AD:
again the source has moved outside the diamond caustic and two images
have merged and disappeared, leaving two images nearly colinear with
the center of the lens.

The four quad and two double image-morphologies are all distinct, and
all the doubles and quads in the CASTLe survey \citep{castles} are
easily classified according to morphology.  But that is only the first
step.  We now show that even without resorting to any modeling,
examination of the relative locations of the images with respect to
each other and to the center of the lensing galaxy can reveal a wealth
of information.

\subsection{The time-ordering of images}\label{time-order}

In doubles, the image further from the galaxy center is a minimum and
arrives first, while the image closer to the galaxy center is a saddle
point and arrives second.  We do not know if this is mathematically
compulsory (intuitively we feel not).  Nevertheless we know of no
exceptions, either observed or in any plausible model.

For quads, time ordering is more subtle.  Extrapolating from doubles,
one might expect that images arrive in order of decreasing distance
from the lens center---but this is not always true: the CLASS lens
B1608+656 (the only quad so far to have its image-ordering observed
directly) refutes this \citep{fassnacht}.  Nevertheless, the idea of
saddle-point contours lets us reliably infer image-ordering in nearly
all cases, including B1608.

\begin{figure}
\epsscale{.4}
\plotone{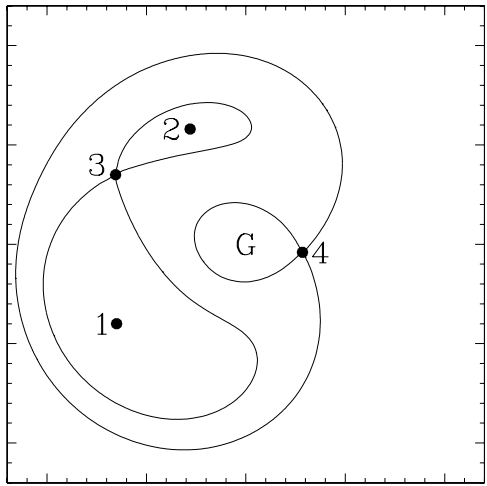}
\caption{Saddle point contours in a generic quad. The images are labeled
in order of their arrival times (Section~\ref{time-order}); images \1 and 
\2 are minima, \3 and \4 are saddle points; the fifth image would be a
maximum at the galaxy's center G. In a generic double, the saddle
point \3 and the minimum \2 are absent, so \4 becomes `\2'.}
\label{fig-sad}
\end{figure}

In Figure \ref{fig-sad} we show saddle point contours in a generic
quad, with images labeled by increasing arrival time.  The two loops
of the lemniscate enclose two minima, the larger loop containing the
deeper minimum \1\ and the smaller loop containing \2.  Of the two
saddle points, the lemniscate has \3\ and the lima\c con has \4.  The
small loop of the lima\c con has the incipient fifth image, which is a
maximum at the center of the galaxy.  (In a double, the lemniscate is
absent; \1\ remains \1, while \4\ becomes \2.)  We can restate the
above as two rules: (i)~\1\ and \2\ are minima, while \3\ and \4\ are
saddle points, and (ii) minima and saddle points alternate in position
angle (PA).  Considering isochrones on the arrival-time surface, we
can formulate two more rules: (iii)~being the global minimum, \1 has
the greatest number of isochrones separating it from the lens center,
so \1\ is the furthest or nearly the furthest from the lens center,
(iv)~analogously \4\ is the closest or nearly the closest to the lens
center.  Finally, examining the merging images in
Figures~\ref{fig-inc} to \ref{fig-short} we see that (v)~if two images
are close together, they will be \2\ and \3.  After a little practice
at using rules (i)--(v), the image ordering is usually obvious.  In
the most symmetric central quads \1\ and \2\ are hard to distinguish,
but apart from this, first impressions and detailed modeling (and
observations, where available) always give the same image order.

We suggest that labeling images by time order would be much
preferable to labeling A, B, C, etc., by brightness as is usual at
present.  First, time order is physically motivated and unique;
brightness order can change with wavelength, and also with time due to
microlensing.  Second, time order is closely connected to the image
configuration, and thus easier to remember.

\subsection{The direction of the source's displacement}\label{source}

From Figures \ref{fig-inc} to \ref{fig-short} it is clear that the PA
of the source with respect to the center of the lensing galaxy
is close to the PA of image \1, which is not surprising
since the global minimum \1\ is the image least changed from the
unlensed case.  Examining these figures further, we can see that the
source does not have exactly the same PA as \1: it always leans away from 
\1 and towards the \2-\3 pair, though by different angles. 
The radial displacement of the source may be very different
from that of the images.

By itself, the location of a source is not very exciting.  However, if
a source has several components, the inferred source position reveals
something of the source morphology. In Section~\ref{casestudies} we
will discuss the example of B1933+507, where three components of a jet
are lensed; there are many lens models, but they all require the
source components to be colinear.

\subsection{The orientation of the main source of shear}\label{shear}

In a circularly symmetric case all the images are equidistant from the
center, forming an Einstein ring. When the source moves off axis, or
the lens becomes moderately elongated, the images still stay
approximately equidistant from the center, typically to within 20\%.
In fact, it is very hard to induce the images to spread over a larger
radial range with one moderately elliptical lensing galaxy.  So a
narrow spread in images' galacto-centric distances indicates small or
zero external shear and moderate galaxy ellipticity.

An external mass (such as another group galaxy) has a different and
easily noticeable effect.  Suppose there is an external mass to the N.
It will contribute most to the arrival time in the N; we can imagine
the external mass pushing up the northern part of the arrival-time
surface.  The effect on the stationary points will be to push the
northernmost and southernmost image towards the center and the
easternmost and westernmost images away from the center; there will be
a large spread in galacto-centric radii among the images.  We have
already come to a similar conclusion via a more algebraic argument
above (near Equation \ref{ecc-ellip}).  Specifically, if there is an
external mass to the N, the long axis of the lens potential will be
NS, and the images will lie on an eccentric ellipse whose major axis
is roughly EW.

Examining the quads in Figures \ref{fig-inc} to \ref{fig-short} again
we can easily find the direction of the shear to within
20--30$^\circ$, just by looking at the images. Detailed modeling does
not do much better then that.

\subsection{The morphology of rings}\label{rings}

Rings and partial rings in multiple-image systems are images of the QSO
host galaxy.  We can predict their appearance approximately, simply
from the arrival-time surface of QSO source. Now, points of zero gradient
on the arrival-time surface correspond to QSO images; we may hope that
regions of small gradient correspond to images of the host galaxy. In
fact, if the host galaxy is modeled by a conical brightness profile,
where the source brightness falls off linearly with distance away 
from the source center until it becomes zero, then
the density of isochrones on the arrival-time surface
represents the inverse brightness of the rings \citep{sw01}.  In other
words, to a first approximation, the ring will be brightest where the
contours on the arrival-time surface are furthest apart.

From the above, we conclude that rings in quads will be brightest
in the arc \1--\2--\3--\4, and fainter or absent in the arc \4--\1
(because arrival-time isochrones will be more closely packed in the
latter region).

Similar conclusions can also be reached by a very different route, via
caustics and critical curves \citep{bk88}.

\section{Model-dependent features}

In the previous section we saw that multiply-imaged lenses permit a
preliminary reconstruction without any modeling; in particular, points
1--4 of the Introduction can be inferred.  Yet some features, in
particular point~0 of the Introduction, are very difficult to
constrain even with detailed modeling.  What is so different about
features like point~0?

Surprisingly, it turns out the properties of an image system
that are shared with its circularly symmetric `cousin' are very
model-dependent and hard to constrain, whereas properties that {\it
arise\/} because of the breaking of circular symmetry are relatively
model-independent and robustly reproduced by even a rudimentary
model. The second category includes points 1--4 above, while point 0
belongs to the first category. The reason can be appreciated
intuitively.  In a circularly symmetric system the mass of the galaxy,
its radial density distribution, and the observer/lens/source distances are
all degenerate; an infinite set of combinations of these parameters
will give rise to exactly the same arrival time surface.  When the
symmetry is broken the arrival time surface acquires new features:
distinct images appear, ring breaks up into pieces, etc. However, the
ambiguity in the galaxy radial profile persists when the symmetry is
broken.

The most important example of a degeneracy inherited from a
circularly-symmetric lens is the magnification transformation
\citep{fgs85} or the mass-disk degeneracy, which can be interpreted as
stretching the arrival-time surface without moving its stationary
points \citep{s00}.  Mathematically, it involves multiplying
$1-\kappa$, $\bbeta$, and $T_0$ (Eq.~\ref{eqT0}), all by the same
constant, $s$. Physically, the degeneracy involves first rescaling the 
lens-mass distribution by $s$, and then adding a mass sheet of surface 
mass density $1-s$, expressed in terms of critical density for lensing.
Rescaling $1-\kappa$ changes the radial density slope of the 
new mass distribution (galaxy+mass sheet), rescaling
$\bbeta$ changes all the magnifications in step (while keeping the
relative magnifications fixed), and rescaling $T_0$ changes the
inferred $h$ from the same data.  The image plane is unaffected except
for the total magnification. Thus $h$ is quite unconstrained by the
modeling unless restrictive assumptions are made.

Aside from the slope of the lens mass density profile, which is a global 
model-dependent feature, there are also local stochastic features,
namely substructure. Substructure of the type that is believed
to exist in and near lens galaxies, and along the lines of sight,
for example, spiral arms, globular clusters, dwarf satellite galaxies, etc.
can modify flux ratios of images somewhat, but will have little effect 
on the positions of images, and virtually no effect on the relative
time delays of images. Thus our conclusions are mostly unaltered by 
substructure.

\section{Case studies}\label{casestudies}

We now show how the well-known quads illustrate the theoretical ideas
of the previous sections.  
The systems we consider are listed in Table~\ref{tbl}. 
We refer to them by the first four digits in their names.
The references to the discovery papers, observational data, and other
information on these systems can be obtained from the CASTLeS Web site,
maintained by \citet{castles}.

In order to illustrate the concepts developed in this paper we replace
the quads with some simple models.  These models are tailored to fit
the image positions reasonably well but {\it not\/} to astrometric
accuracy.  The models are used to obtain the shape of critical and
caustic curves, and the saddle point contour in the arrival time
surface. The qualitative features 1--4 (cf.~Section~\ref{intro}) are
just as well reproduced by our simple models as by more sophisticated
ones.

As model potential we use an elliptical isothermal with external shear
\begin{equation}
\psi = (a\theta_x^2+b\theta_y^2+2h\theta_x\theta_y+r_{\rm c}^2)^{1/2} +
       \gamma_1 \half (\theta_x^2-\theta_y^2) +
       \gamma_2 \theta_x \theta_y.
\label{modpot}
\end{equation}
The core radius $r_{\rm c}$ is fixed at $0.1$, while the other parameters
were determined by modeling; their values are listed in Table~\ref{tbl}. 
The values of the apparent shear, $\gamma_{\rm app}$, introduced in 
Section~\ref{appshear} are also listed in the table. 
 
\begin{figure}
\epsscale{.55}
\plotone{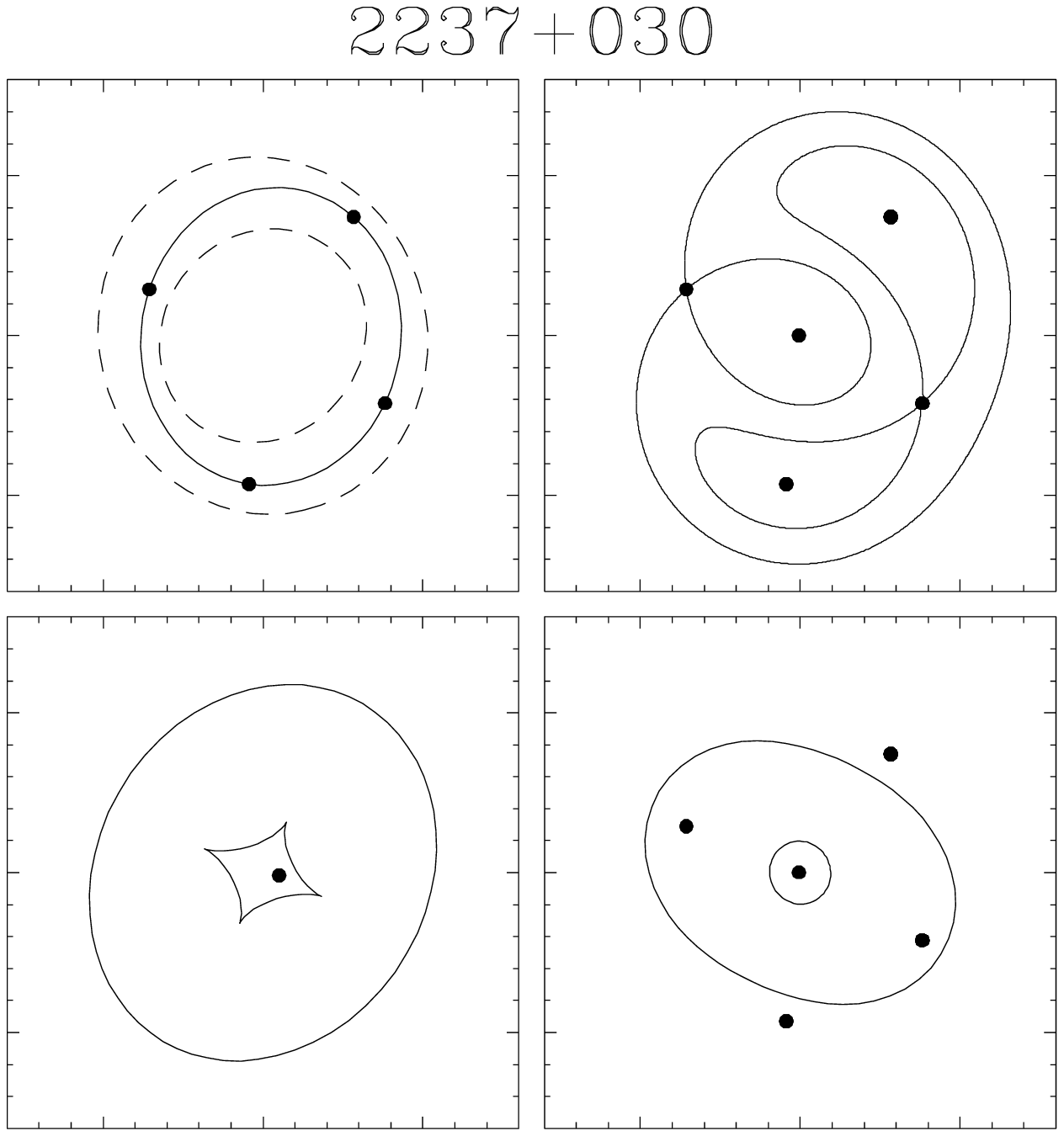}
\caption{Model for 2237, a canonical \CQ. Three of the panels are
standard (and again the caustic panel is expanded 2-fold).  In the
upper-left panel, the solid curve is the apparent-shear ellipse 
(Section~\ref{appshear}), the
inner dashed ellipse shows the axes ratio of the lens potential, while
the outer dashed ellipse shows the external shear. The sizes of the
dashed ellipses are arbitrary, only the axis ratio and orientation is
relevant. Note that the image locations are from the simple parametric model
of Equation~\ref{modpot}, and not the actual observed image positions.
The differences between the two are small, and unimportant for our present
purposes. 2237, and other observed quads are discussed in 
Section~\ref{casestudies}.}
\label{fig-2237}
\end{figure}

\begin{figure}
\epsscale{.55}
\plotone{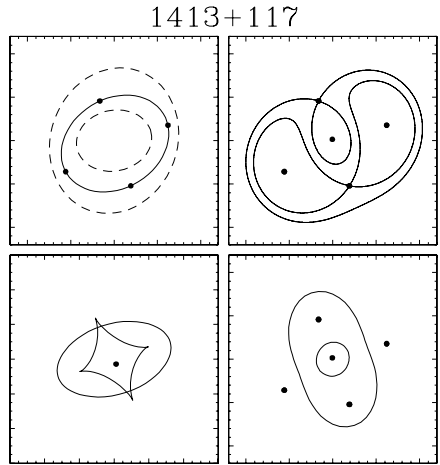}
\caption{Model for 1413, another canonical \CQ. Panels follow the same
plan as in Figure~\ref{fig-2237}.}
\label{fig-1413}
\end{figure}

\begin{figure}
\epsscale{.55}
\plotone{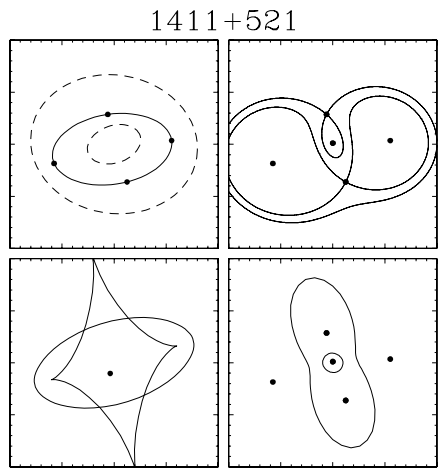}
\caption{Same as Figure~\ref{fig-2237}, but for 1411, a \CQ.}
\label{fig-1411}
\end{figure}

2237, 1413, and 1411 (Figs.~\ref{fig-2237}, \ref{fig-1413}, and
\ref{fig-1411} respectively) are all \CQ s, because their four images
are approximately evenly distributed around the lens center. In the
first two lenses, images \1 and \2 are almost equidistant from the
center, so it is hard to distinguish them using rule (iii) of
Section~\ref{time-order}; but this is the only ambiguity. Note,
however, that when \1 and \2 are equidistant from lens center, which
one is the global minimum becomes a moot point. 1413 is one of the most
symmetric of all known quads, but even in this case we can apply rule
(iv) and identify the upper left image as \4. The saddle of the
lemniscate, \3, is then the lower right image, by rule (i). 2237 is
less symmetric; after identifying the upper left image as \4, we
notice that the two lower images are the closest pair in the system,
making them \2 and \3, by rule (v).  Thus the upper right image is the
global minimum, \1.  The position of the source with respect to the
lens center can be determined approximately by looking at the PA of
\1. 2237 and 1413 do not have very large shears. On the other hand,
1411 appears to have a very elliptical lens, which is reflected in the
large value of $\gamma_{\rm app}$.  If the source were located
closer to the top
of the plot, we could have had a lemniscate double rather than a lima\c con
double---common in cluster lenses, but not in galaxy lenses. 1411 is
ideal for illustrating the effects of an elliptical lens, described in
Section~\ref{shear}.  The upper and lower images are visibly closer to the
lens center compared to the right and left images, which indicates
that the galaxy's mass distribution is elongated along the upper-left to
lower-right direction.  The upper image, being closest to the center, is
\4. The galaxy's elongation is evident from the size and orientation of
the diamond caustic; thus modeling confirms qualitative conclusions
based on image geometry.  \1 and \2 are hard to distinguish,
suggesting that both minima are of nearly equal depth, though the
left image is probably \1. 

\begin{figure}
\epsscale{.55}
\plotone{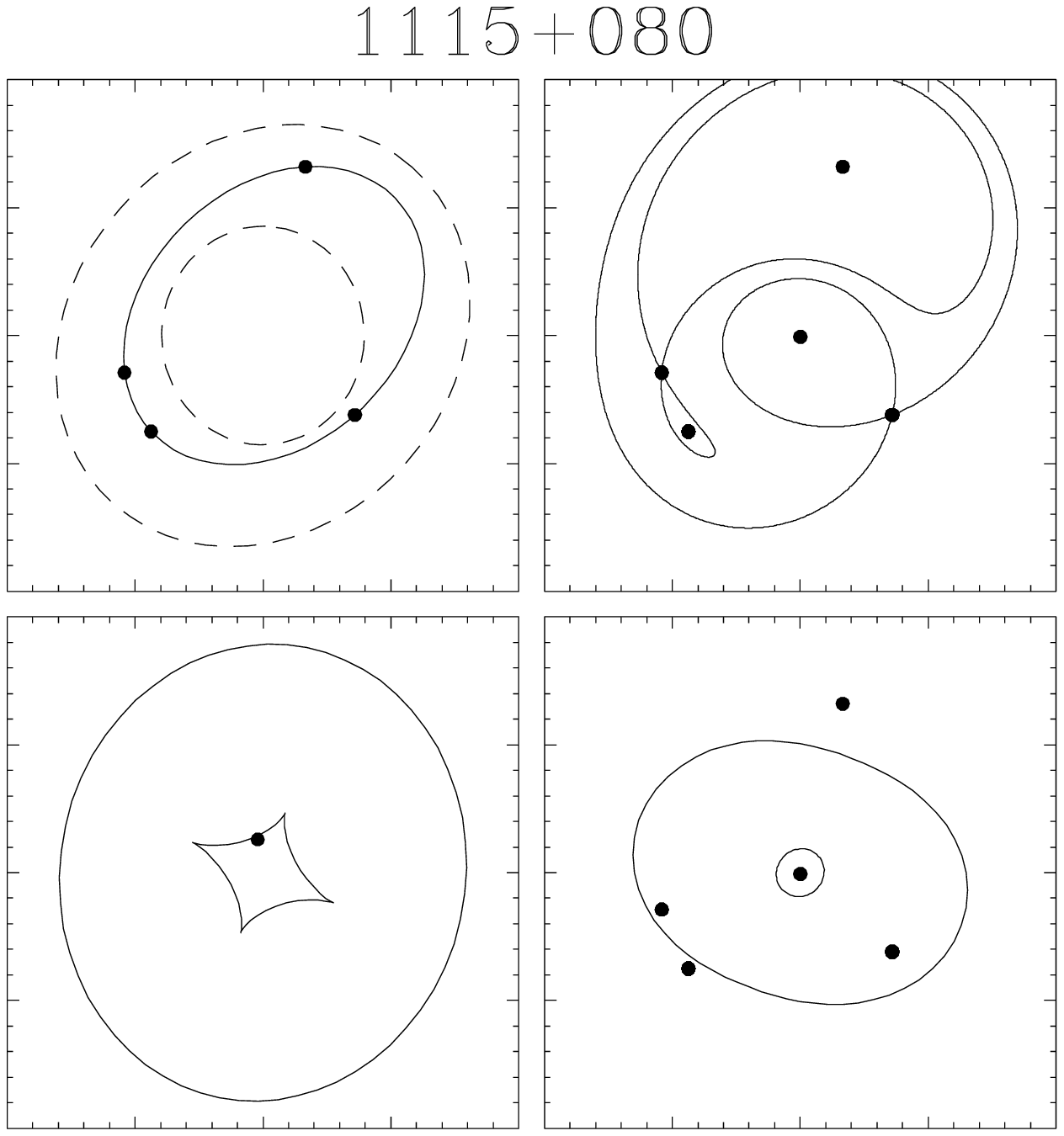}
\caption{Same as Figure~\ref{fig-2237}, but for 1115, an \IQ.}
\label{fig-1115}
\end{figure}

\begin{figure}
\epsscale{.55}
\plotone{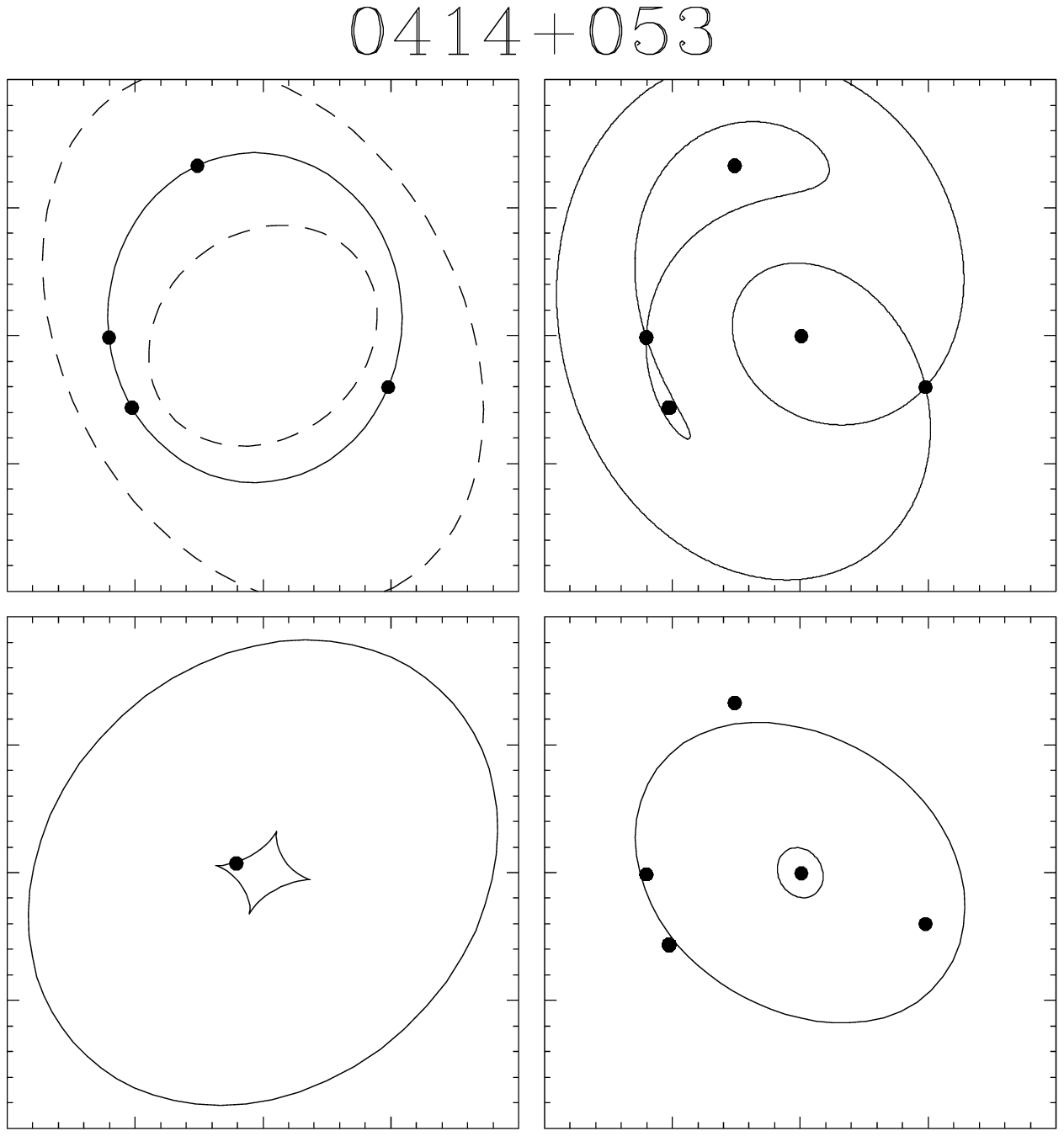}
\caption{Same as Figure~\ref{fig-2237}, but for 0414, an \IQ.}
\label{fig-0414}
\end{figure}

\begin{figure}
\epsscale{.55}
\plotone{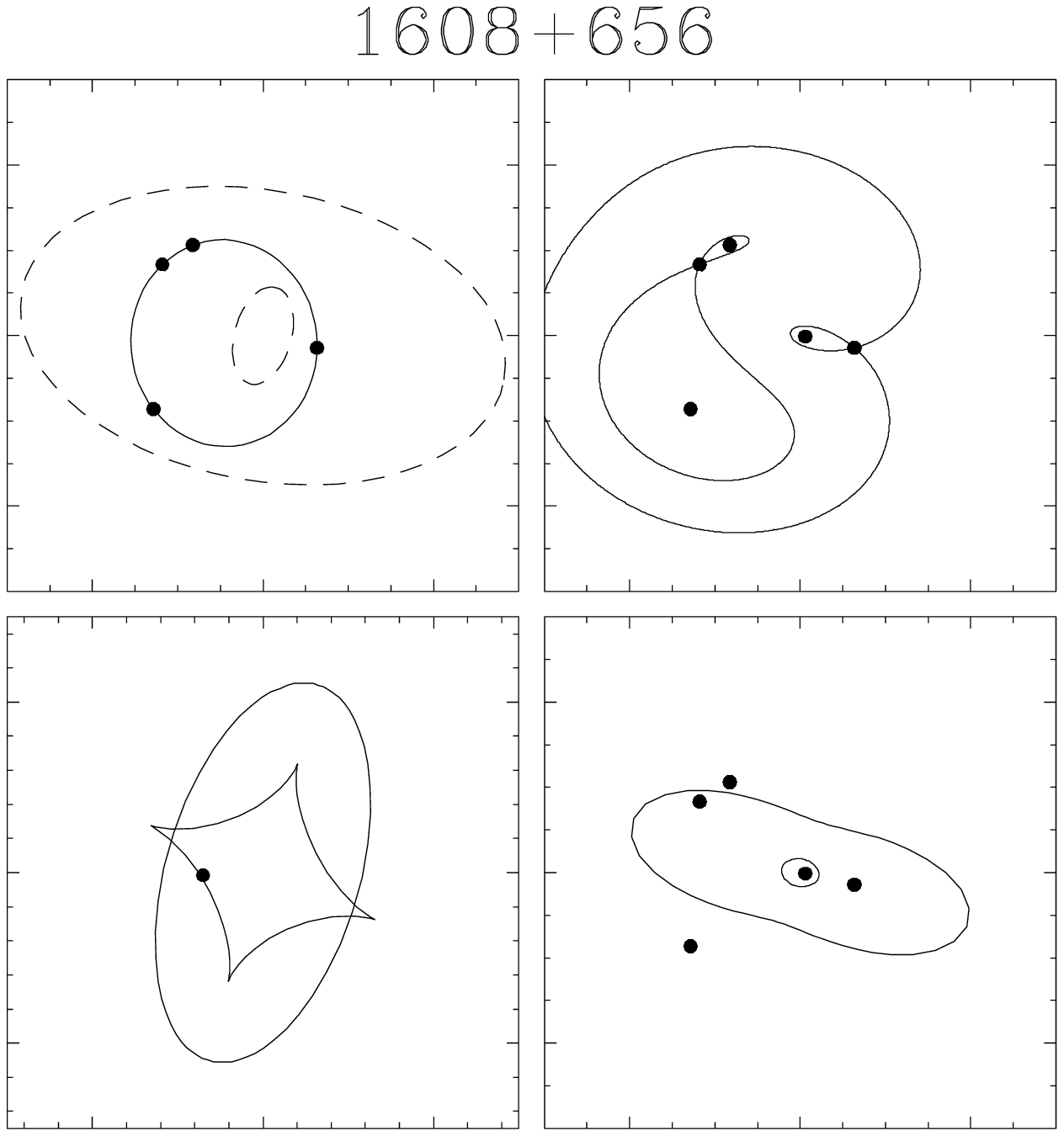}
\caption{Same as Figure~\ref{fig-2237}, but for 1608, an \IQ.
Because we used a simple lens potential (Equation~\ref{modpot}), our model 
for 1608 is strictly speaking inaccurate, though it is still 
instructive---see text.}
\label{fig-1608}
\end{figure}

1115, 0414, and 1608 (Figs.~\ref{fig-1115}, \ref{fig-0414}, and
\ref{fig-1608} respectively) are \IQ s.  Such quads are the easiest to
time order.  The upper images in 1115 and 0414 and the lower image in
1608 are \1. The most identifiable feature of \IQ s is the close pair
of images, \2 and \3 according to rule (v).  The rest follows from
rules of Section~\ref{time-order}.  In all three cases time ordering
yields the pattern of extrema which is also revealed by the saddle
point contours of Figures~\ref{fig-1115}, ~\ref{fig-0414}, and
~\ref{fig-1608}.

1115, which has a source of external shear in the form of the group, 
demonstrates points 3 and 0 of the Introduction, that the direction of 
shear is robustly recovered by any modeling, while the value of $h$ is
much more model-dependent. \citet{kk97}, using an SIS ellipsoid and a modified 
Hubble profiles for the galaxy mass distribution obtained best fit position 
angles of the group of $-125^\circ$ and $-118^\circ$, respectively. 
Using a completely different type of modeling, non-parametric pixellated
lens, \citet{sw97} found that the source of shear is located at $-130^\circ$
(see their Fig.~2). This $\sim 5-10\%$ precision in determining PA of shear 
is to be contrasted with $\geqsim 40\%$ uncertainty in the final $h$.

1115 has a well known ring, made up of merging images of QSO's extended
host galaxy. The faintest segment is the arc \4--\1 \citep{impey98}, 
in agreement with the qualitative predictions of Section~\ref{rings}.

Comparison of 0414 and 1608 shows the effect of shear in \IQ s. While
0414's images are approximately equidistant from center, implying
small shear, 1608's \4 is much closer to the center than the other
images, indicating the presence of shear, whose axis of mass
distribution is nearly horizontal.  However, 1608 is more complex than
that: shear arises because the galaxy-lens is a double, possibly an
interacting pair, and our simple model potential does not allow for
such a complicated lens.  Instead, 1608 is fit by a very elliptical
potential (responsible for moving \4 closer to center, and \3 further
away from center) with an opposite external shear (responsible for
orienting \1,\3,\2 images horizontally across from \4).

1608 is pushing approximations to (and sometimes beyond) their
limits. For example, \citet{wmk00}, assuming a relatively general
parametric lens model, derive a relation between image galacto-centric
radii and time delays (their Eq.~[12]).  Applying this to 1608's
images \2 and \3 would actually result in negative $h$ (because image
\3 is further than \2 from the lens center).  Adding significant
external shear (their Eq.~[18]) can produce positive $h$, but still
not realistic values. 

\begin{figure}
\epsscale{.55}
\plotone{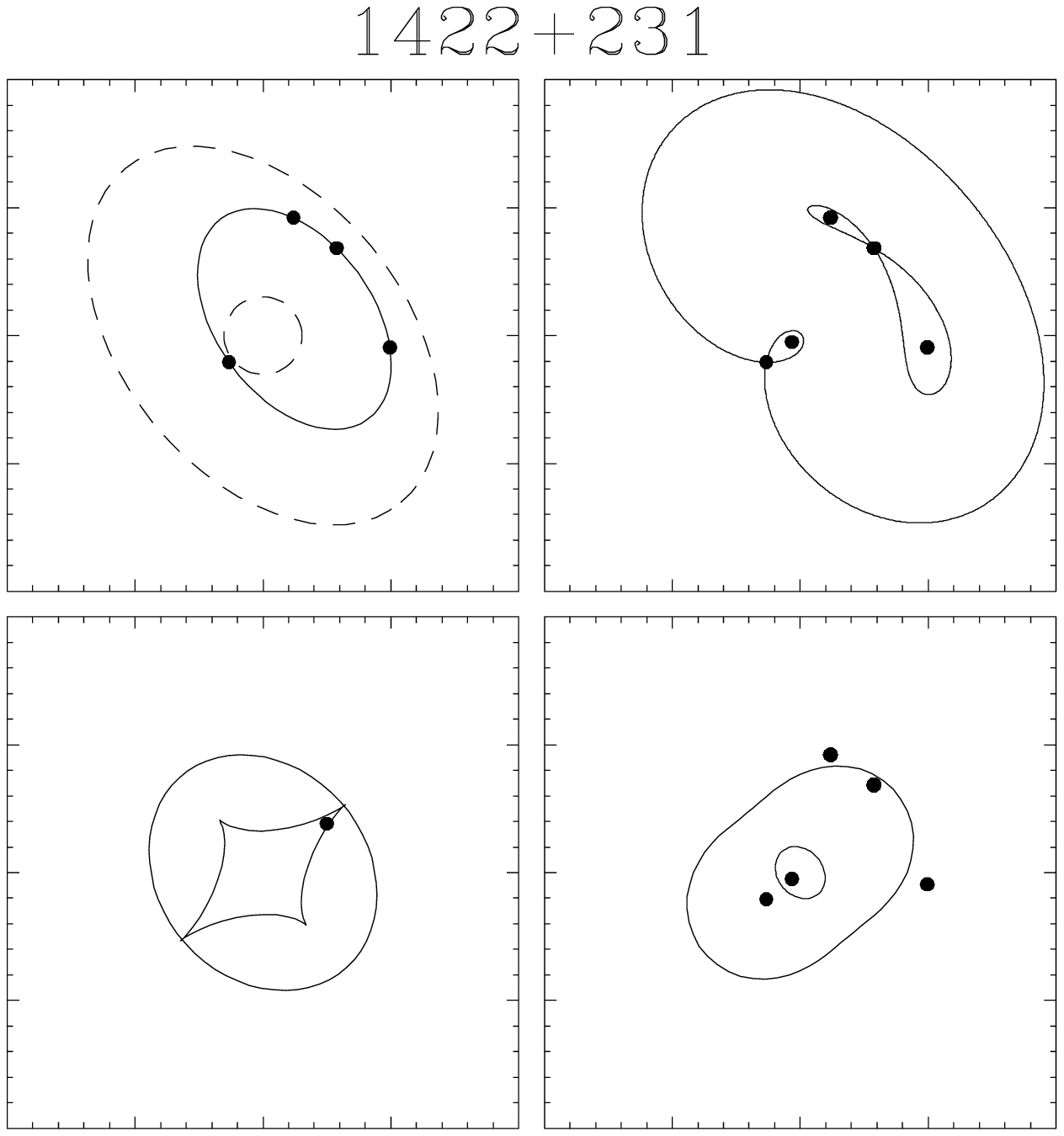}
\caption{Same as Figure~\ref{fig-2237}, but for 1422, a \LQ.}
\label{fig-1422}
\end{figure}

\begin{figure}
\epsscale{.55}
\plotone{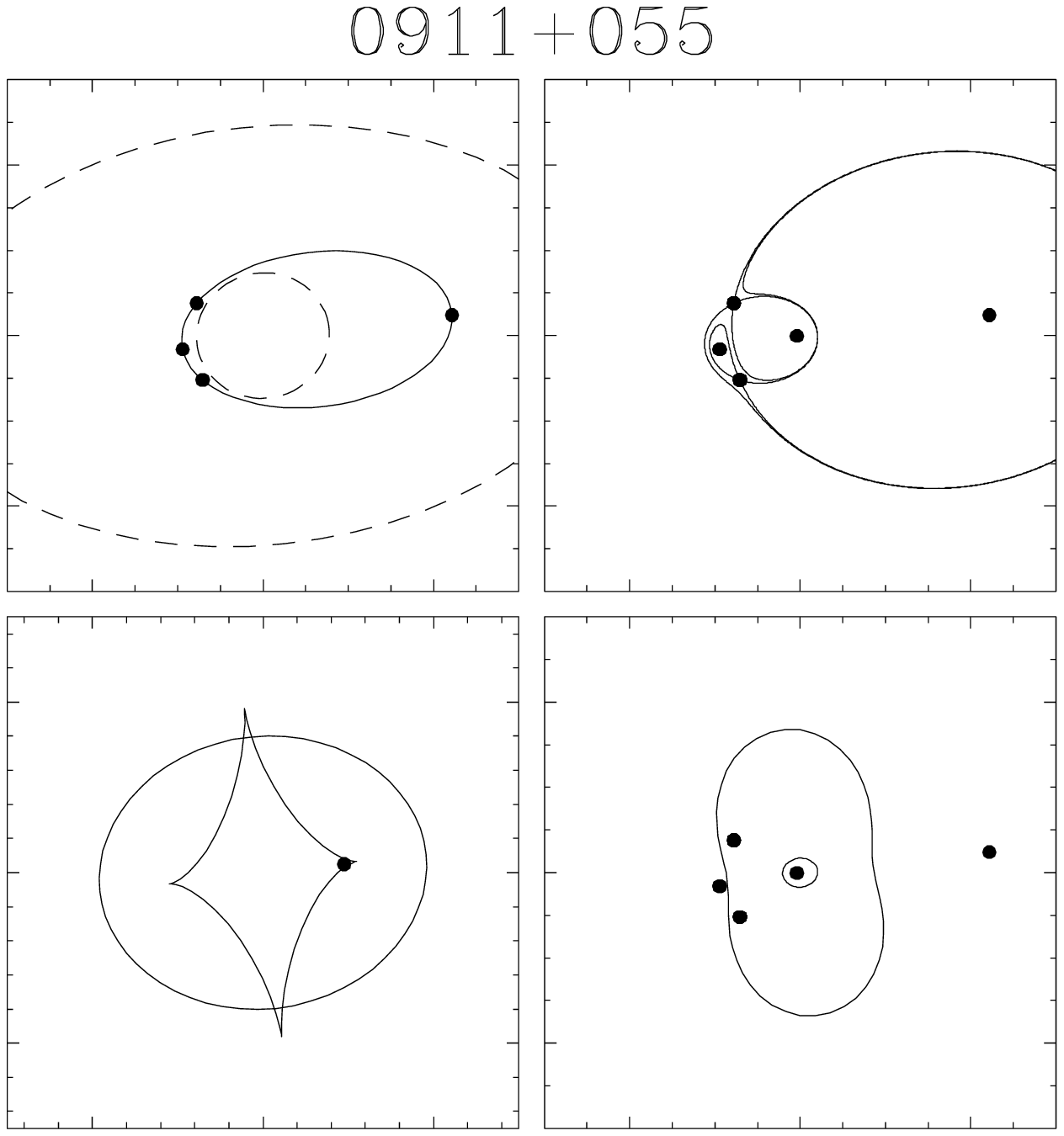}
\caption{Same as Figure~\ref{fig-2237}, but for 0911, a \SQ.}
\label{fig-0911}
\end{figure}

1422 and 0911 (Figs.~\ref{fig-1422} and \ref{fig-0911}) have large
external shear due to neighboring groups \citep{burud98,kun97} as
indicated by the distances of images from lens centers. Morphology
tells us that for 1422 the source is along the long axis while for
0911 source is along the short axis (see upper panels of
Figs.~\ref{fig-long} and ~\ref{fig-short}).  In \SQ s, such as 0911,
the direction towards \1 is a good indicator of the direction of the
source away from the lens center.  Image ordering is
straightforward---rule (v) helps here: the closest pair is \2 and \3.
1422 and 0911 are textbook examples of the systems where external
shear dominates and the source is located on the same axis as the
galaxy-lens and group (1422), and on an axis perpendicular to the
galaxy-group axis (0911). 

The cluster-lens Cl0024+1654 is another \LQ.

\begin{figure}
\epsscale{.55}
\plotone{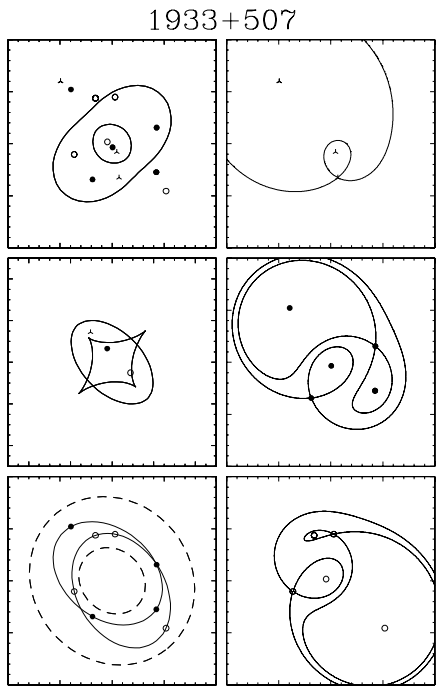}
\caption{Model for 1933, a combination of three multiple-image
systems; filled circles denote the images and sources of a \CQ, open
circles refer to an \IQ, and skeletal stars refer to an \ID.  The
right-hand panels show the arrival-time surface for each system.  The
upper-left panel show the critical curves and images; note the double
hook traced out by the images.  The middle-left panel shows caustics
and source positions. The lower-left panel shows apparent-shear
ellipses for both quads as solid curves. The
inner dashed ellipse shows the axes ratio of the lens potential, while
the outer dashed ellipse shows the external shear.}
\label{fig-1933}
\end{figure}

1933 is a 10 image system; it contains a \CQ, an \IQ, and an \ID\
(Fig.~\ref{fig-1933}).  The double-hook traced by images indicates
that sources lie nearly in a line.  To see this, recall that for a
centrally concentrated lens, the image of a line (i.e., the solution
of one Cartesian component of the lens equation) is a line with a kink
or double-hook (e.g. Witt 1993).  Applying the rules of
Section~\ref{source} to the two quads and one double of this system
(see the three right-hand panels of Fig.~\ref{fig-1933}) one can
infer that the three sources are nearly in a straight line.  Judging
by their radio spectral indices and morphology, the three sources are
probably the core of a radio galaxy flanked by two radio features,
possibly jet components \citep{nair}.

\section{Exceptional systems}\label{complex}

The ideas we have discussed in this paper suffice to gain a good
qualitative understanding of nearly all multiply-imaged QSOs.  But
Nature is cunning, and very occasionally a lens turns up that looks
like nothing else.  Such exceptions occur if the source or the lens
has multiple compact components.

Multiple source components can be disentangled using spectral
properties: in the case of 1933 (discussed above) the different source
components have different spectral steepness in radio, and hence the
ten images can be unambiguously grouped into two quads and a double.

Systems with multiple lens components are more difficult to figure
out.  We comment briefly on two such lenses: B1359+154, discovered and
interpreted by \citet{rusin01}, and PMN J0134-0931, discovered by
\citet{winn02} and interpreted by \cite{kw03}.

The six-image system 1359 is very nicely summarized in Fig.~6 of
\citet{rusin01}, especially panel~c.  From their figure, we
immediately recognize the morphology of a long-axis quad, akin to
1422.  We can even guess from the figure that external shear will be
contributed by galaxies (roughly) to the WNW or ESE, which is
consistent with the detailed results of their paper.  But---and this
is the unique feature---the central potential well is due not to one
galaxy but to a compact group of three galaxies. Consequently, what
would have been the demagnified maximum of a lima\c con has been split
by two internal lemniscates, yielding two new saddle points (observed)
and three maxima (demagnified and unobserved).

The five-image system 0134 is even more curious.  The lens evidently
consists of two galaxies (possibly a merging pair) producing three
saddle points and two minima.  One might perhaps describe it as a
distant relative of a long-axis quad, but really the morphology is
completely unlike any other known object.

We look forward to more developments on 0134 and other exceptional
systems that expose the limitations of our work. But meanwhile, we
hope researchers will find the simple ideas of this paper helpful in
understanding the remaining (vast majority of) multiply-imaged QSOs!

\acknowledgments

This has been a somewhat unusual paper and produced some lively
discussion during the peer-review process.  We thank the two referees
and the Editor for their comments; even where we disagreed it all
helped improve the paper.

\appendix
\section{Interactive Java applet}

\begin{figure}
\epsscale{0.95}
\plotone{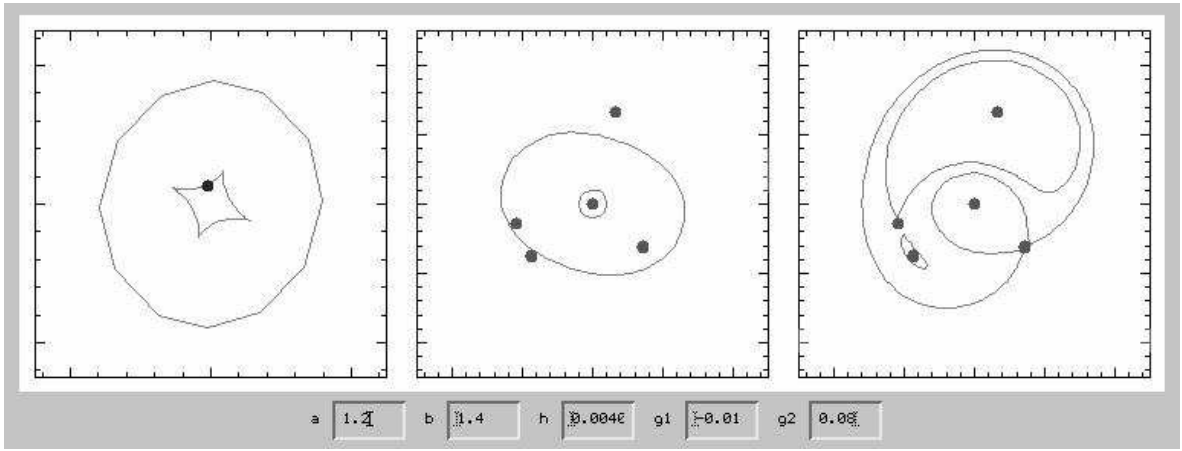}
\caption{Java applet showing caustics, critical curves, and saddle-point
contours. The user types in values for $a$, $b$, $h$, $\gamma_{\rm 1}$ and 
$\gamma_{\rm 2}$ (see Eq.~\ref{modpot}) in the text field at the bottom
of the applet. Source position is chosen by clicking in the left panel.} 
\label{fig-applet}
\end{figure}

We have developed a Java applet to show caustics, critical curves, and
saddle-point contours interactively.  It is illustrated in Figure
\ref{fig-applet}.  The user types in parameter values for the model
potential (\ref{modpot}) in the text fields, and inputs a source
position by clicking the mouse.  The applet then computes and plots
the invisible curves and the image positions.

The source code for this and related Java applets are available from
the authors' web pages.


\begin{table}
\begin{center}
\caption{Parameters in the model potential (\ref{modpot}) and the
apparent shear (defined in subsection \ref{appshear}) for some
well-known systems.  The ten-image system 1933 includes two quads and
hence has two apparent-shear values.}
\begin{tabular}{crrrrrll}
\noalign{\null\bigskip}
\tableline\tableline
\noalign{\smallskip}
system & \multicolumn1c{$a$} & \multicolumn1c{$b$} & \multicolumn1c{$h$}
       & \multicolumn1c{$\gamma_1$} & \multicolumn1c{$\gamma_2$}
       & \multicolumn1c{$\gamma_{\rm app}$} \\
\noalign{\smallskip}
\tableline
\noalign{\smallskip}
Q2237+030    & $ 0.750$  & $ 0.800$  & $ 0.075$  & $-0.040$  & $-0.020$ & $0.070$ \\
H1413+117    & $ 0.450$  & $ 0.300$  & $ 0.033$  & $-0.060$  & $ 0.050$ & $0.118$ \\
HST14113+5211& $ 1.100$  & $ 0.600$  & $ 0.188$  & $ 0.090$  & $-0.030$ & $0.263$ \\
PG1115+080   & $ 1.200$  & $ 1.400$  & $ 0.004$  & $-0.010$  & $ 0.080$ & $0.138$ \\
MG0414+0534  & $ 1.500$  & $ 1.400$  & $ 0.251$  & $-0.100$  & $-0.140$ & $0.058$ \\
B1608+656    & $ 0.800$  & $ 2.050$  & $ 0.369$  & $ 0.240$  & $-0.090$ & $0.054$ \\
B1422+231    & $ 0.600$  & $ 0.600$  & $ 0.000$  & $-0.040$  & $-0.210$ & $0.225$ \\
RXJ0911+0551 & $ 1.450$  & $ 1.300$  & $ 0.027$  & $ 0.240$  & $ 0.050$ & $0.274$ \\
B1933+507    & $ 0.290$  & $ 0.290$  & $-0.040$  & $-0.010$  & $-0.050$ &
                                             \multicolumn2l{$0.158\pm 0.007$} \\
\noalign{\smallskip}
\tableline
\end{tabular}
\label{tbl}
\end{center}
\end{table}

\end{document}